\documentclass[twocolumn]{jpsj2}

\usepackage{graphicx}% Include figure files
\usepackage{dcolumn}% Align table columns on decimal point
\usepackage{bm}% bold math

\title{Metal-insulator transition in the two-orbital Hubbard model 
at fractional band fillings: 
Self-energy functional approach}

\author{Kensuke {\sc Inaba}$^1$ and
Akihisa {\sc Koga}$^2$}
\inst{%
$^1$ Department of Applied Physics, Osaka University, Suita, Osaka 565-0871 , Japan\\
$^2$ Department of Physics, Kyoto University, Kyoto 606-8502, Japan
}

\recdate{\today}% It is always \today, today,
\abst{
We investigate the infinite-dimensional two-orbital Hubbard model 
at arbitrary band fillings. By means of the self-energy functional approach,
we discuss the stability of the metallic state 
in the systems with same and different bandwidths.
It is found that the Mott insulating phases are realized 
at commensurate band fillings. 
Furthermore, it is clarified that
the orbital selective Mott phase with one orbital localized and 
the other itinerant is stabilized even at fractional band fillings
in the system with different bandwidths.
}

\kword{self-energy functional approach, orbital selective Mott transition}
\begin{document}
\maketitle

\section{Introduction}
Recently, strongly correlated electron systems with orbital degeneracy
have attracted much interest.\cite{ImadaRev}
Among them, heavy fermion behavior in the transition metal oxides
is one of the most important issues in the condensed matter physics.
In the lithium vanadate $\rm LiV_2O_4$,\cite{Kondo97}
geometrical frustration originating from 
the pyrochlore structure suppresses 
magnetic correlations, 
realizing a heavy fermion state below $T \sim 40$K.\cite{Kaps01,Isoda00} 
The orbital degeneracy in the $3d$ bands was also suggested 
to stabilize the heavy fermion state.\cite{Fujimoto01,Tsunetsugu02,Yamashita03} 
Other interesting examples are the transition metal oxides 
$\rm Ca_{2-x}Sr_xRuO_4$\cite{Nakatsuji} and $\rm La_{n+1}Ni_nO_{3n+1}$.
\cite{Kobayashi96,LaNiO}
It is suggested that a Mott transition in some of relevant orbitals,
{\it i.e.} an orbital selective Mott transition (OSMT),\cite{Anisimov02}
is induced by the chemical substitution of Ca ions 
and also by the change in temperatures, 
where the heavy metallic state is realized.
Furthermore, the pressure-induced OSMT 
is also suggested in the vanadium oxide $\rm V_2O_3$,\cite{Laad06}
which stimulates further theoretical
\cite{Liebsch03,Liebsch03L,Liebsch04,Sigrist04,Okamoto04,Fang04,Koga04,Koga05,Koga05Pr,KogaRev,Ferrero05,Medici05,Arita05,Knecht05,Liebsch05,Biermann05,Tomio05,Inaba05os,Ruegg05,Inaba06,Dongen06,Werner06} 
and experimental \cite{Balicas05,Lee06} investigations 
on the effect of electron correlations 
in the system with orbital degeneracy.

The two-band system in infinite dimensions is one of the simplest models
with orbital degeneracy,
where many groups have discussed
%the effect of electron correlations,
%\cite{Gunnarsson96,Han98,Ono03}
the nature of the Mott transition,
\cite{Bunemann:Gutzwiller,Gunnarsson96,Kotliar96,Rozenberg97,Hasegawa98,Klejnberg98,Han98,Imai01,Koga:ED_DMFT,Florens02,Florens02b,Ono03,Pruschke05} 
the magnetism,\cite{Momoi98,Held98,Sakai06} %etc. 
the thermodynamic properties,\cite{Oudovenko02,Inaba05} etc.
Among them, the possibility of the OSMT related to the above compounds
has been discussed recently.\cite{Anisimov02,Fang04,Sigrist04,Okamoto04,Laad06}
It has been clarified that 
separate Mott transitions occur in the half-filled system 
with different bandwidths, in general.\cite{Koga04}
On the other hand, in a certain parameter region, 
the effect of the different bandwidths is diminished, where
a single Mott transition is induced.\cite{Liebsch03L,Koga04,Medici05,Ferrero05} 
In contrast to these intensive studies of the bandwidth control 
Mott transitions,
little is known about the filling control Mott transitions,
which may be important to discuss the stability of the OSMT 
in real materials.
Therefore, it is desired to discuss 
the filling control Mott transitions
as well as the bandwidth control Mott transitions systematically.

Our concern is to investigate the two-orbital Hubbard model 
at arbitrary band fillings.
For this purpose, we make use of the self-energy functional approach (SFA),
\cite{Potthoff03a,Potthoff03b,Pozgajcic04}
which allows us to discuss the Mott transition systematically.
By calculating several physical quantities, 
we discuss the stability of the metallic state in the system.

This paper is organized as follows.
In Sec. \ref{sec_model_method},
we introduce the two-orbital Hubbard model.
In Sec. \ref{sec_detail},
we treat the two-orbital Hubbard model with same bandwidths
to test the validity of our analysis.
We discuss the nature of the Mott transition in the system 
with different bandwidths to determine the phase diagram
in Sec. \ref{sec_diffband}.
A brief summary is given in Sec. \ref{sec_summary}
%%%%%%%%%%%%%%%%%%%%%%%%%%%%%%%%%%%%%%%%%%%%%%%%%%%%%%%%%%%%%%%%%%%%%
%%%                        Model and Method                        %%
%%%%%%%%%%%%%%%%%%%%%%%%%%%%%%%%%%%%%%%%%%%%%%%%%%%%%%%%%%%%%%%%%%%%%
\section{Model and Method}\label{sec_model_method}
%====================================================================
%==                              model                             ==
%====================================================================
We investigate the two-orbital Hubbard model with different bandwidths,
which is explicitly given as,
${\cal H}={\cal H}_0+{\cal H}^\prime$,
${\cal H}'=\sum_i {\cal H}_i^\prime$ with
%%%%%%%%%%%%%%%%%%%%%%%%%%%%%%%
\begin{eqnarray}
{\cal H}_0&=&\sum_{<i,j>, \alpha, \sigma}%\sum_{\alpha}\sum_{\sigma}
\left( t_\alpha -\mu\delta_{ij}\right)
          c^\dag_{i\alpha\sigma} c_{j\alpha\sigma},\\
%{\cal H}^\prime&=&\sum_i{\cal H}^{\prime}(i)\nonumber\\
%{\cal H}^{\prime}(i)&=&
{\cal H}_i^\prime&=& \!\!\!\!
                  U \sum_{\alpha}
                n_{i \alpha \uparrow} n_{i \alpha \downarrow}
                  +\sum_{\sigma \sigma^\prime}
                 (U^\prime - \delta_{\sigma\sigma^\prime}J)
                n_{i 1 \sigma} n_{i 2 \sigma^\prime}
                \nonumber\\
          &-&  \!\!\!\!\! J  (c^\dag_{i 1 \uparrow}c_{i 1 \downarrow}
                c^\dag_{i 2 \downarrow}c_{i 2 \uparrow}
                +c^\dag_{i 1 \uparrow}c^\dag_{i 1 \downarrow}
                c_{i 2 \uparrow}c_{i 2 \downarrow}+H.c.)
          \label{eq_model},
\end{eqnarray}
%%%%%%%%%%%%%%%%%%%%%%%%%%%%%%%%%%%%%%%%%%
where $c^\dag_{i\alpha\sigma}(c_{i\alpha\sigma})$ 
is a creation (annihilation) operator
of an electron with spin $\sigma (=\uparrow, \downarrow)$ and orbital
$\alpha (=1,2)$ at the $i$th site, 
and $n_{i\alpha\sigma}$ is the number operator.
Here, $t_\alpha$ denotes the hopping integral for the $\alpha$th orbital, 
$\mu$ the chemical potential, 
$U (U')$ the intra-orbital (inter-orbital) Coulomb interaction, 
and $J$ the Hund coupling including the spin-flip and pair-hopping terms. 
We impose the condition $U=U'+2J$, 
which is obtained by the symmetry arguments for the degenerate orbitals.

%====================================================================
%==                             Method                             ==
%====================================================================
To discuss the competition between the metallic and the Mott insulating phases,
we make use of the SFA proposed recently.
\cite{Potthoff03a,Potthoff03b}
Since this method is based on the variational principle,
it has an advantage in discussing the nature of the Mott transitions.
In fact, it has been applied to correlated electron systems 
at half filling, where the precise phase diagrams have been obtained.
\cite{Potthoff03a,Potthoff03b,Pozgajcic04,Inaba05,Inaba05os}
Here, we deal with the two-orbital system at arbitrary band fillings
to determine the phase diagrams.
The detail of the SFA for the doped system is explicitly shown in Appendix.
In the paper, we use a semi-circular density of states (DOS),
$\rho_\alpha(\omega)=4/\pi W_\alpha \sqrt{1-(2\omega/W_\alpha)^2}$,
where $W_\alpha$ is a bandwidth for the $\alpha$th orbital.

In the next section,
we treat the degenerate Hubbard model with same bandwidths
as a simple model, to check the validity of our analysis for the doped system.
Then we determine the phase diagram in the system 
to discuss the role of the Hund coupling for the Mott transitions
at quarter filling.

%%%%%%%%%%%%%%%%%%%%%%%%%%%%%%%%%%%%%%%%%%%%%%%%%%%%%%%%%%%%%%%%%%%%%
%%%                 the system with same bandwidth                 %%
%%%%%%%%%%%%%%%%%%%%%%%%%%%%%%%%%%%%%%%%%%%%%%%%%%%%%%%%%%%%%%%%%%%%%
\section{
Mott transitions in the system with same bandwidths}\label{sec_detail}

We consider the two-orbital Hubbard model with same bandwidths 
$(W_1=W_2=4)$.
%%%%%%%%%%%%%%%%%%%%%%%%%    rnm_fill      %%%%%%%%%%%%%%%%%%%%%%%%
\begin{figure}[htb]
\includegraphics[width=\linewidth]{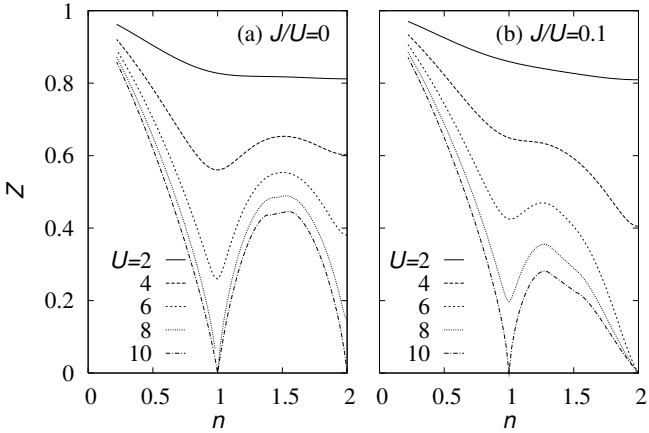}
\caption{Renormalization factor as a function of the total electron count
$n$ in the case $J/U=0$ (a) and $J/U=0.1$ (b).
}\label{fig:rnm_fill}
\end{figure}
%%%%%%%%%%%%%%%%%%%%%%%%%%%%%%%%%%%%%%%%%%%%%%%%%%%%%%%%%%%%%%%%%%%%%%%%%%%
The renormalization factor $Z$, 
which is proportional to the inverse of the effective mass,
is shown in Fig. \ref{fig:rnm_fill}.
First, we focus on the case $J/U=0$, as shown in Fig. \ref{fig:rnm_fill} (a). 
The introduction of the Coulomb interaction monotonically decreases 
the renormalization factor and
the double dip structure appears in the case $(U>4)$.
In the strong coupling region, 
the renormalization factor reaches zero when $n=1$ and $n=2$, 
while it never vanishes otherwise.\cite{Rozenberg97,Hasegawa98}
This result is consistent with the well-known fact that 
the Mott transition occurs only at 
commensurate band fillings such as quarter filling $(n=1)$ 
and half filling $(n=2)$. \cite{Rozenberg97,Hasegawa98,Ono03}
The corresponding critical points $U_c \sim 9.2$ and $7.7$ are 
in agreement with those obtained by other numerical methods.
\cite{Koga:ED_DMFT,Ono03}
Therefore, we can say that the reliable results are obtained in terms of the SFA
not only at half filling but also at arbitrary band fillings.
A similar dip structure appears in the system with a finite Hund coupling 
$J/U=0.1$, as shown in Fig. \ref{fig:rnm_fill} (b).
%This suggests that
%the Mott transitions
%occur at commensurate band fillings, as similar to the case $J=0$.

\begin{figure}[htb]
\includegraphics[width=\linewidth]{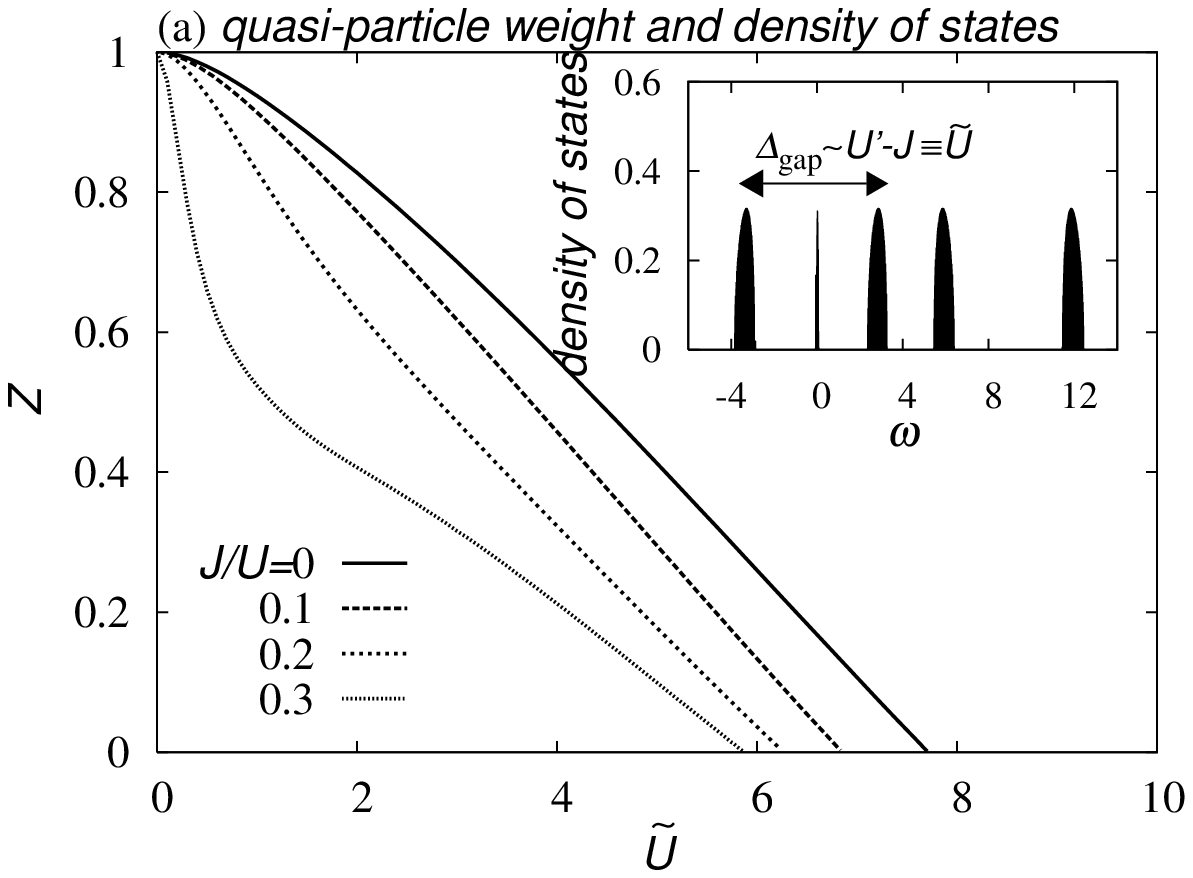}
\includegraphics[width=\linewidth]{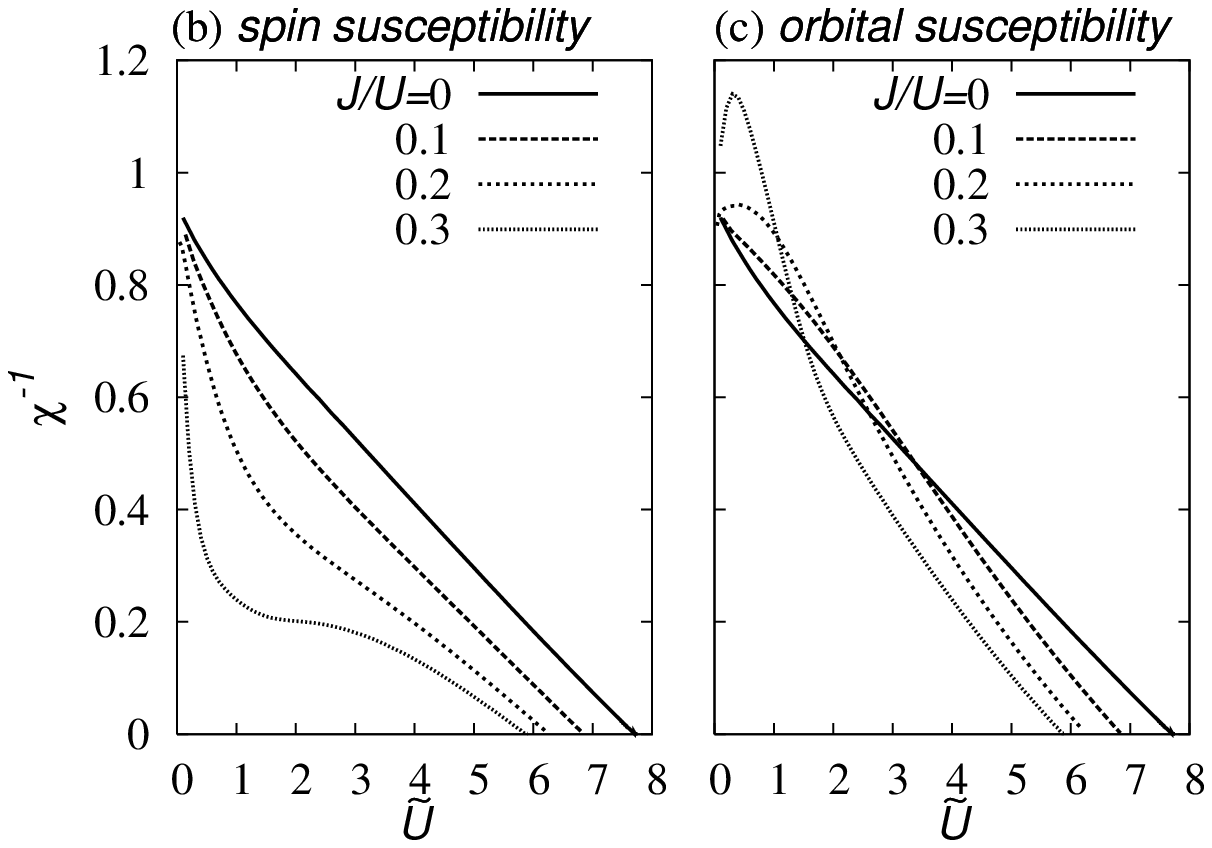}
\caption{
The quasi-particle weight (a), the spin susceptibility (b),
the orbital susceptibility (c) as a function of $\tilde{U}$ 
when $n=1$.
The inset of the panel (a) is the density of state $-{\rm Im} G_\alpha(\omega)/\pi$
for $J/U=0.2$, $U=15$ and $U'=12$, 
where both the renormalized metallic state at Fermi level and
the precursor of the Hubbard satellites appear.
}\label{fig_rnm_J}
\end{figure}

We now address how the Hund coupling affects the competition between 
the metallic and the Mott insulating phases
at quarter filling $(n=1)$,
in comparison with the results for the half filled case $(n=2)$.
\cite{Ono03,Pruschke05,Inaba05}
In Fig. \ref{fig_rnm_J}, we show the renormalization factor,
and the spin and orbital susceptibilities.
In the case, the parameter $\tilde{U}=U'-J$
is useful to discuss the Mott transition in the system,
which corresponds to the energy scale of the lowest Hubbard gap
[see the inset of Fig. \ref{fig_rnm_J} (a)].
%%%%%%%%%%%%%%%%%%%%%%%%%    fig_rnm_J      %%%%%%%%%%%%%%%%%%%%%%%%
%%%%%%%%%%%%%%%%%%%%%%%%%%%%%%%%%%%%%%%%%%%%%%%%%%%%%%%%%%%%%%%%%%%%
When $J/U$ is small, 
the increase of the interaction $\tilde{U}$
decreases the renormalization factor, where 
spin and orbital susceptibilities are monotonically enhanced.
%%%%%%%%%%%%%%%%%%%%   dos J  %%%%%%%%%%%%%%%%%%%%%
\begin{figure}[htb]
\includegraphics[width=\linewidth]{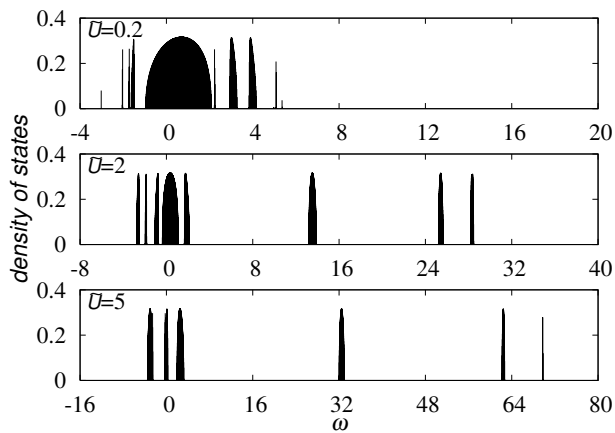}
\caption{
The density of states for $J/U=0.3$ with several choices of $\tilde{U}=0.2$, $2$, and $5$. 
}\label{fig_dos_J}
\end{figure}
%%%%%%%%%%%%%%%%%%%%%%%%%%%%%%%%%%%%%%%%%%%%%%%%%%%%%%%%%%%%%%%%%%%%
On the other hand, the large Hund coupling $(J/U=0.3)$ 
leads to different behavior.
It is found that 
the introduction of $\tilde{U}$ rapidly decreases the renormalization factor,
where the spin (orbital) susceptibility is increased (decreased).
%We show the density of states with several choices of $\tilde{U}$
%at Fig. \ref{fig_dos_J}.
When %$\tilde{U}=0.2$ and
%since the condition 
$\tilde{U}=U'-J \lesssim W \ll U, U'$ ($\tilde{U}=0.2$), %is satisfied,
%the interaction $\tilde{U}$ does not induce the Hubbard bands,
the Hubbard gap does not appear near the Fermi level,
but high-energy structures appear due to $U$ and $U'$,
as shown in the top panel of Fig. \ref{fig_dos_J}.
%in comparison with the strength of the interactions $U$ and $U'$, 
%$\tilde{U}(=U'-J)$ is negligible in the region.
%Therefore,
It suggests that
the spin-triplet two-electron state formed by $J$
plays an important role in low energy states,
%---------------------------added-------------------------
while the spin-singlet states contribute the high energy features.
%while the heavy fermion state renormalized by $U$ and $U'$ 
%appears around the Fermi level.
However, when the system approaches the Mott critical point (the bottom panel of Fig. \ref{fig_dos_J}),
%---------------before
%%%  all the Hubbard satellites are shifted to the high energy region. 
%---------------after
the Hubbard satellite that originates from the spin-triplet state appears at the position of $\tilde{U}$,
while the Hubbard satellites have roots in the spin-singlet states are shifted to higher energy region.
%as shown in middle and top panels of Fig. \ref{fig_dos_J}.
Then, the triplet state does not directly  
contribute to low energy states,
%where an energy cost of 
but contributes 
through a virtual process between the ground state and the triplet state.
%which is a first excited state at the atomic limit
%becomes relevant in the low energy excitation.
Therefore, the heavy fermion state is gradually changed in character
around $\tilde{U}\sim 0.3$.
Eventually, both susceptibilities simultaneously diverge 
at a critical point $\tilde{U}_c=5.9$ $(U_c=58.7)$, where
the second-order transition occurs to the Mott insulating phase, 
as shown in Fig. \ref{fig_rnm_J}.

\begin{figure}[htb]
\includegraphics[width=\linewidth]{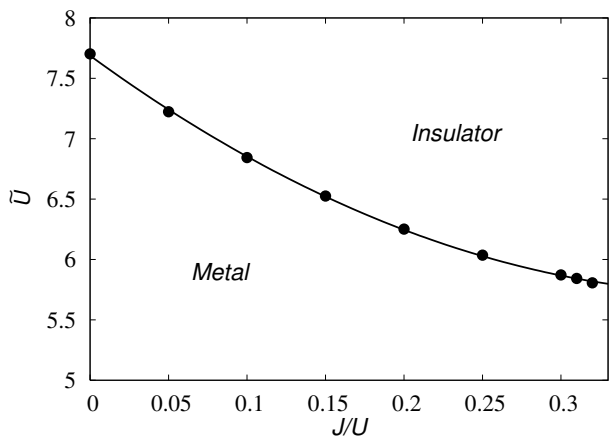}
\caption{
The phase diagram for the system with same bandwidths at quarter filling.
%The phase boundary is a guide to the eye.
}\label{fig_Uc_J}
\end{figure}

By performing similar calculations, we end up with the phase diagram 
as shown in Fig. \ref{fig_Uc_J}.
%%%%%%%%%%%%%%%%%%%%%%%%%%%    fig_Uc_J    %%%%%%%%%%%%%%%%%%%%%%%%%
%%%%%%%%%%%%%%%%%%%%%%%%%%%%%%%%%%%%%%%%%%%%%%%%%%%%%%%%%%%%%%%%%%%%%%%%%%%
Note that at quarter filling, spin and orbital fluctuations are simultaneously 
enhanced at the critical point, yielding the second-order Mott transition.
This behavior is in contrast to that for the half filled case.
\cite{Koga:ED_DMFT,Ono03,Inaba05}
At half filling, the second-order Mott transition occurs in the case $U=0$, 
where spin and orbital fluctuations are enhanced simultaneously. 
On the other hand, if the Hund coupling is introduced, orbital fluctuations
are strongly suppressed, yielding the first-order Mott transition.
\cite{Bunemann:Gutzwiller,Ono03,Inaba06}

In this section, we have studied the degenerate Hubbard model 
with same bandwidths by means of the SFA.
The obtained results reproduce the well-known numerical results,
\cite{Rozenberg97,Hasegawa98,Ono03}
implying that the SFA allows us to investigate 
strongly correlated electron systems 
with and without particle-hole symmetry systematically.
In the following, we consider the two-orbital system with different bandwidths
to discuss ground state properties
at arbitrary band fillings.

%%%%%%%%%%%%%%%%%%%%%%%%%%%%%%%%%%%%%%%%%%%%%%%%%%%%%%%%%%%%%%%%%%%%%
%%%              the system with different bandwidths              %%
%%%%%%%%%%%%%%%%%%%%%%%%%%%%%%%%%%%%%%%%%%%%%%%%%%%%%%%%%%%%%%%%%%%%%
\section{The system with different bandwidths}\label{sec_diffband}

Let us move our attention to the effect of different bandwidths. 
Recently, theoretical advances have been made 
in the system at half filling, 
where the nature of the OSMT has been discussed.
\cite{Liebsch03L,Koga04,Arita05,Biermann05,Medici05,Ferrero05,Ruegg05,Knecht05,Werner06}
It has been clarified that separate Mott transitions occur in general, 
which merge to a single Mott transition under a certain condition.
\cite{Koga04,Liebsch03L,Medici05,Ferrero05}
Furthermore, it has been discussed how stable the OSM phase is against
the hole doping,\cite{Koga04} the hybridization,\cite{Koga05} 
the magnitude and/or the anisotropy of the Hund coupling,\cite{Koga04,Liebsch04,Inaba05os} etc.
However, it is not trivial whether 
the OSM phase in the doped system is adiabatically connected to
the Mott insulating phase at quarter filling.
This may be important to understand real materials, in which 
the total electron count can be tuned by the chemical substitutions.
Therefore, it is necessary to systematically discuss  how 
the Mott insulating phases, which are realized in commensurate band fillings,
are stabilized in the doped system, which
may be a key to understand the Mott transitions in real materials
with orbital degeneracy. 

\begin{figure}[htb]
\includegraphics[width=\linewidth]{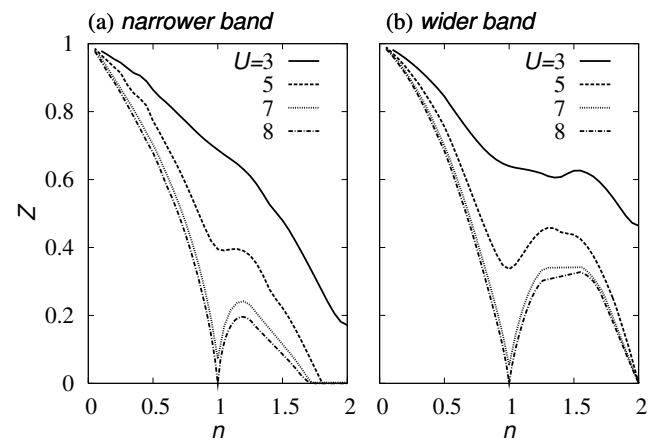}
\caption{Renormalization factor for narrower (wider) orbital (a) [(b)] 
as a function of the total electron count $n$
in the system with $R=0.5$ and $J/U=0.1$}
\label{fig:rnm_R05_J01}
\end{figure}

Here, we consider the Mott transitions in the Hubbard model with
the different bandwidths $W_1=2$ and $W_2=4$ $(R=0.5)$.
%to study the competition between the metallic and Mott insulating phases. 
%Then, we discuss how stable the OSM phase is against the hole doping and 
%electron correlations.
To overview 
the competition between the metallic and the Mott insulating phases,
we first calculate the renormalization factor for each orbital
in the system with $J/U=0.1$, as shown in Fig. \ref{fig:rnm_R05_J01}.
%%%%%%%%%%%%%%%%%%%%%%%%%    fig_rnm_BW    %%%%%%%%%%%%%%%%%%%%%%%%
%%%%%%%%%%%%%%%%%%%%%%%%%%%%%%%%%%%%%%%%%%%%%%%%%%%%%%%%%%%%%%%%%%%%
When $n=2$, half filling is realized in each band, where
the effective Coulomb interaction should depend 
on the bandwidth of each orbital $(U_\alpha^{eff}=U/W_\alpha)$. 
Therefore, with the increase of the interaction, 
the renormalization factor for the narrower band are decreased rapidly.
Eventually, the OSMT occurs at the critical point $U_c \sim 3.3$,
inducing the OSM phase with one orbital localized and the other itinerant.
Further increase of interaction induces another Mott transition 
in the wider band
at the critical point $U_c \sim 4.0$.
These results are consistent with the previous results.\cite{Medici05,Inaba05os}
In contrast, somewhat complicated behavior appears away from half filling, where
an electron count for each band depends on the magnitude of the interactions.

\begin{figure}[htb]
\includegraphics[width=\linewidth]{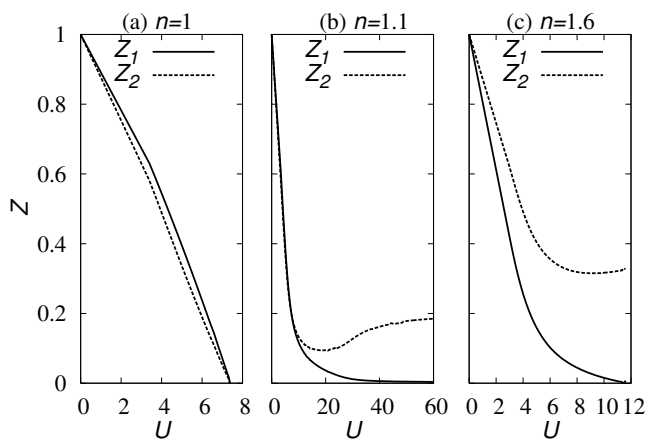}
\caption{The renormalization factor for narrower (wider) orbital (a) [(b)] 
as a function of the Coulomb interaction $U$ 
in the system with $R=0.5$ and $J/U=0.1$}
\label{fig:rnm}
\end{figure}

To make this clear, we calculate the renormalization factor and 
the electron count for each orbital with a fixed total electron count
$n=1.0$ (quarter filling), 1.1 and $n=1.6$,
as shown in Figs. \ref{fig:rnm} (a), (b), and (c).
%%%%%%%%%%%%%%%%%%%%%%%%%    fig_rnm_BW    %%%%%%%%%%%%%%%%%%%%%%%%
%%%%%%%%%%%%%%%%%%%%%%%%%%%%%%%%%%%%%%%%%%%%%%%%%%%%%%%%%%%%%%%%%%%%
%%%%%%%%%%%%%%%%%%%%%%%%%    fig_rnm_BW    %%%%%%%%%%%%%%%%%%%%%%%%
\begin{figure}[htb]
\includegraphics[width=\linewidth]{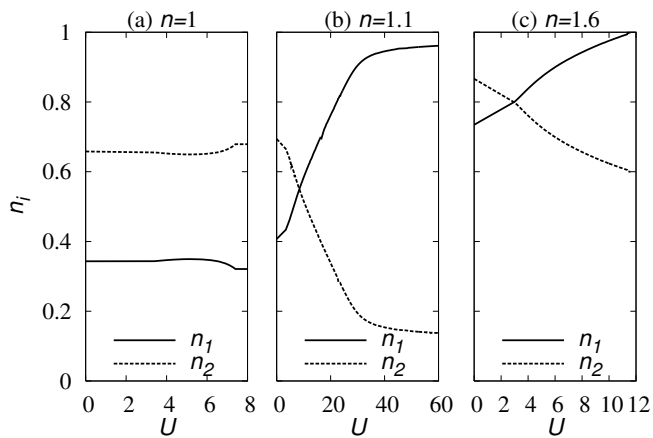}
\caption{The electron count for narrower (wider) orbital (a) [(b)] 
as a function of the Coulomb interaction $U$
in the system with $R=0.5$ and $J/U=0.1$}
\label{fig:fill}
\end{figure}
%%%%%%%%%%%%%%%%%%%%%%%%%%%%%%%%%%%%%%%%%%%%%%%%%%%%%%%%%%%%%%%%%%%%
%When $U=0$, the renormalization factor is unity and 
%the number of electrons for each band is smaller than unity.
Now, we focus on the case $n=1.6$, 
where the total electron count is close to half filling.
As increasing the interactions,
the renormalization factor for the narrower band 
reaches zero at a critical point $U_c=11.7$,
where half filing is realized in the narrower band.
This implies that the commensurability for the narrower band 
is satisfied in the presence of electron correlations, 
and thus the system is driven to the OSM phase.\cite{Koga04}
On the other hand, when the system is close to quarter filling $(n=1.1)$,
the electron count for both bands is far from unity
even in the strong coupling regime, as shown in Fig. \ref{fig:fill} (b).
This implies that 
the heavy quasi-particle state persists 
and the Mott transition never occurs 
(at least, up to $U\sim 60$).

In contrast, at quarter filling $(n=1)$, 
the renormalization factors $z_1$ and $z_2$ reach zero simultaneously, 
where the single transition occurs.
It is found that 
the electron count for each orbital still remains fractional
even in the Mott insulating state.
This suggests that the commensurability is never satisfied
 for either orbital.
An important point is that 
the Mott transition at quarter filling is quite different 
from the OSMT discussed above,
implying that the Mott insulating state is not 
adiabatically connected to the OSM state.
To clarify the difference in these Mott transitions,
it is instructive to discuss spin and orbital fluctuations in the system 
at arbitrary band fillings.

\begin{figure}[htb]
\includegraphics[width=\linewidth]{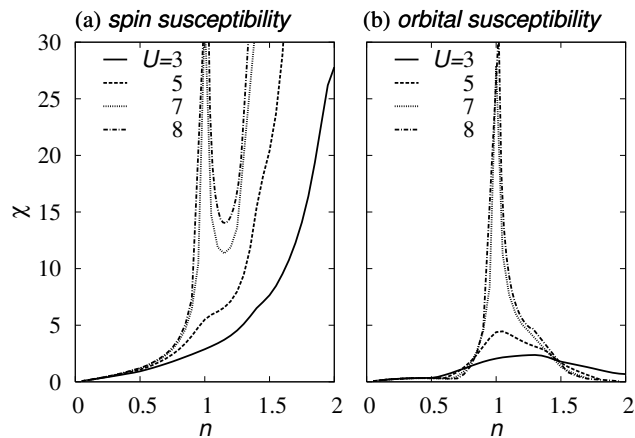}
\caption{Spin (a) and orbital (b) 
susceptibilities as a function of the total electron 
count $n$ when $J/U=0.1$.}
\label{fig:sus}
\end{figure}

In Fig. \ref{fig:sus}, we show the local susceptibilities for spin and 
orbital degrees of freedom.
%%%%%%%%%%%%%%%%%%%%%%%%%    fig_sus    %%%%%%%%%%%%%%%%%%%%%%%%
%%%%%%%%%%%%%%%%%%%%%%%%%%%%%%%%%%%%%%%%%%%%%%%%%%%%%%%%%%%%%%%%%%%%
Characteristic behavior of the Mott transition appears in the figure.
When $n\sim 1$, the increase of the interaction enhances
spin and orbital fluctuations simultaneously, inducing the Mott transition 
at quarter filling.
On the other hand, when the system is close to half filling $(n \gtrsim 1.3)$,
the spin susceptibility is enhanced and the orbital susceptibility 
is strongly suppressed
in the strong coupling region, 
where the OSM phase is realized with one orbital localized and 
the other itinerant.
An important point is that 
the metallic state between these Mott insulating phases
are quite sensitive to the total electron count, as discussed above.
Therefore, it is expected that the ground state is drastically changed 
by some perturbations, {\it e.g.} the crystalline electric field, 
the chemical substitution, etc.

\begin{figure}[htb]
\includegraphics[width=\linewidth]{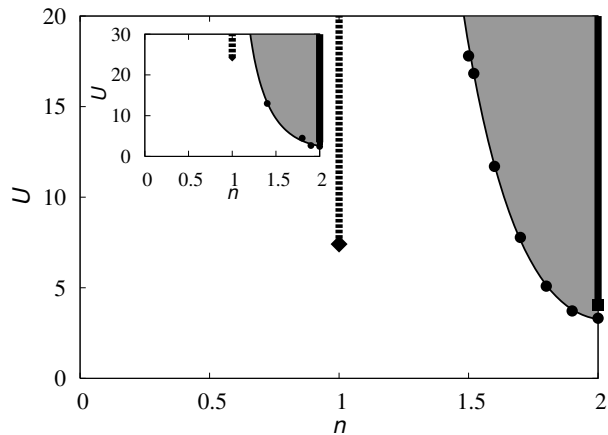}
\caption{Phase diagram for the system with different bandwidths $R=0.5$ 
when $J/U=0.1$ ({\it main}) and $J/U=0.25$ ({\it inset}). 
The shaded region represents the OSM phase 
with one orbital localized and the other itinerant.}
\label{fig:phase}
\end{figure}

We end up with the phase diagram as shown in Fig. \ref{fig:phase}.
%%%%%%%%%%%%%%%%%%%%%%%%%    fig_rnm_BW    %%%%%%%%%%%%%%%%%%%%%%%%
%%%%%%%%%%%%%%%%%%%%%%%%%%%%%%%%%%%%%%%%%%%%%%%%%%%%%%%%%%%%%%%%%%%%
At half filling $(n=2)$, as introducing the Coulomb interaction,
the Mott transition in the narrower band occurs to the OSM phase.
Further increase of the interaction drives the system 
to the Mott insulating phase.
If the total electron count is changed in the Mott insulating phase, 
holes are doped only in the wider band due to the strong correlations 
in the narrower band.
Therefore, the OSM phase is stabilized even away from half filling. 
When the electron count approaches quarter filling, 
the holes are introduced in both bands, 
thus the system is driven to the metallic state. 
In contrast, the different type of the Mott insulating state appears
beyond a certain critical point when the system is quarter-filled.

Up to now, we have restricted our discussion 
to the condition $J/U=0.1$ and $R=0.5$.
If the Hund coupling is increased, orbital fluctuations are further suppressed.
Therefore, it is expected that the OSM phase becomes more stable, while
the Mott insulating phase at quarter filling becomes unstable.
In fact, in comparison with the phase diagram for $J/U=0.1$ and $R=0.5$,
the region of the OSM phase broadens and
the critical point at quarter filling is increased, 
as shown in the inset of Fig. \ref{fig:phase}.
%At that time, the correlated metallic region shrinks on the increase of 
%the Hund coupling, implying that the metallic state is quite sensitive to 
%other perturbations. 
The large difference of the bandwidths 
also suppresses orbital fluctuations, as discussed in the previous paper.\cite{Inaba06}
Therefore, %it is expected that
the same phase diagram is expected to be obtained,
where the wide OSM phase away from half filling is realized.
%On the other hand, when $J/U\rightarrow 0$,
%orbital fluctuations are enhanced, where 
%the effect of different bandwidths is diminished and 
%the single Mott transition occurs at half filling.
%Therefore, the phase diagram similar to that for the system 
%with same bandwidths is obtained (which is not shown in the paper).

%%%%%%%%%%%%%%%%%%%%%%%%%%%%%%%%%%%%%%%%%%%%%%%%%%%%%%%%%%%%%%%%%%%%
%%%                            Summary                            %%
%%%%%%%%%%%%%%%%%%%%%%%%%%%%%%%%%%%%%%%%%%%%%%%%%%%%%%%%%%%%%%%%%%%%
\section{Summary} \label{sec_summary}

We have investigated the two-orbital Hubbard model with same and 
different bandwidths at zero temperature.
By making use of the SFA, we have discussed 
the competition between the metallic and the insulating states
in the system at arbitrary band fillings.
We have found that the Mott transition occurs only at commensurate band 
fillings in the system with same bandwidths.
On the other hand, in the system with different bandwidths, 
the OSM phase with one orbital localized and the other itinerant is stabilized
in a certain parameter region. 
There are some open problems
% could not be addressed 
in the present study.
One of the most important issues is magnetism of the system, 
which has not been treated here, 
since we have restricted our attention to the 
paramagnetic phase. This problem is under 
consideration.

\section*{acknowledgments}
%\begin{acknowledgments}
We would like to thank S. Suga, N. Kawakami, M. Sigrist, T. M. Rice 
and A. Liebsch for valuable discussions.
Numerical computations were carried out at the Supercomputer Center, 
the Institute for Solid State Physics, University of Tokyo. 
This work was supported by a Grant-in-Aid for Scientific Research from 
the Ministry of Education, Culture, Sports, Science, and Technology, Japan.
%\end{acknowledgments}

%%%%%%%%%%%%%%%%%%%%%%%%%%%%%%%%%%%%%%%%%%%%%%%%%%%%%%%%%%%%%%%%%%%%
%%%                       SFA                      %%
%%%%%%%%%%%%%%%%%%%%%%%%%%%%%%%%%%%%%%%%%%%%%%%%%%%%%%%%%%%%%%%%%%%%
\appendix{}

\section{Formulation of the SFA}\label{sec_insulator}

%====================================================================
%==                             Method                             ==
%====================================================================
%In this paper, 
%to investigate the two-orbital Hubbard model at an arbitrary band filling,
%we make use of the SFA proposed recently.
%Since this method is based on the variational principle,
%it is powerful to discuss the nature of the Mott transitions.
%\cite{Potthoff03a,Potthoff03b}
%In fact, this SFA has successfully been applied to strongly correlated 
%electron systems to discuss the Mott transitions at half filling.
%However, little work has dealt with the doped systems.
%\cite{Aichhorn04,Senechal05,Aichhorn05,Aichhorn05EL}
%When one discusses the Mott transition in the system 
%without particle-hole symmetry, 
%it may be difficult to determine the variational parameters uniquely.
We briefly summarize the formulation of the SFA 
%\cite{Potthoff03a,Potthoff03b}
in order to 
discuss the Mott transition in the system without particle-hole symmetry.
Here, by combining the numerical method with the analytic method,
we describe the Mott insulating state in the framework of the SFA.

First, let us consider the free energy $F$ for the system as,
\begin{equation}
F/L=\Omega/L+\mu n,\label{eq_FreeEnergy}
\end{equation}
where $L$ is the number of sites and $n$ is the electron count per site.
The grand potential $\Omega$ is given as,
%------------------------   grand potential   -----------------------
\begin{eqnarray}
\Omega [{\boldsymbol \Sigma}]&=&\Gamma[{\boldsymbol \Sigma}] + 
{\rm Tr}\ln [-({\bf G}_0^{-1}-{\boldsymbol \Sigma})^{-1}]\label{eq:LW},
\end{eqnarray}
%--------------------------------------------------------------------
where $\Gamma[{\boldsymbol \Sigma}]$ is the Legendre transformation of 
the Luttinger-Ward potential. \cite{Luttinger}
${\bf G}_0$ and ${\boldsymbol \Sigma}$ are the 
bare Green function  and the self-energy, respectively.
The condition 
${ \partial \Omega[{\boldsymbol \Sigma}] }/{ \partial {\boldsymbol \Sigma}} =0$
gives us the Dyson equation ${\bf G}^{-1}={\bf G}_0^{-1} 
-{\boldsymbol \Sigma}$,
\cite{Luttinger}
where ${\bf G}$ is the full Green function. 
Note that the potential $\Gamma[{\boldsymbol \Sigma}]$ does not 
depend on the detail of the non-interacting Hamiltonian.\cite{Potthoff03a}
This enables us to introduce a reference system with the same interacting term.
The Hamiltonian is explicitly given by
${\cal H}_{\rm ref}({\bf t}')={\cal H}_0({\bf t}')+{\cal H}'$ 
with the parameter matrix ${\bf t}'$.
Then the grand potential is given as,
%--------------------------------------------------------------------
\begin{eqnarray}
\Omega [{\boldsymbol \Sigma}({\bf t}^\prime)]&=&\Omega({\bf t^\prime})
\nonumber\\
    &+&{\rm Tr}\ln
    \left[-(\omega+\mu-{\bf t}-{\boldsymbol \Sigma}
({\bf t^\prime}))^{-1}\right]
    \nonumber\\
    &-&{\rm Tr}\ln
    \left[-(\omega+\mu-{\bf t}'-{\boldsymbol \Sigma}
({\bf t^\prime}))^{-1}\right],\label{eq:omega_SFA}
\end{eqnarray}
%--------------------------------------------------------------------
where $\Omega({\bf t}')$ and ${\boldsymbol \Sigma}({\bf t}')$ are 
the grand potential and the self-energy for the reference system.
The condition 
$\partial \Omega[{\boldsymbol \Sigma({\bf t}')}]/\partial {\bf t}'=0$
gives us an appropriate reference system in the framework of the SFA,
which approximately describes the original correlated system.
 
Here,
we introduce a two-site Anderson impurity model as the reference system,
which has successfully described strongly correlated electron systems 
in the infinite dimensions.
\cite{Potthoff03a,Potthoff03b,Balzer05,Koller05,Inaba05,Inaba05os}
The Hamiltonian of the reference system is explicitly given as
%---------------    Hamiltonian of the reference system    --------------
\begin{eqnarray}
{\cal H}_{\rm ref}&=&\sum_i{\cal H}_{\rm ref}^{(i)},\\
%---------------    Hamiltonian of the reference system    --------------
  {\cal H}_{\rm ref}^{(i)}&=&\sum_{\alpha \sigma } \varepsilon^{c}_{\alpha}
          c^\dag_{i\alpha\sigma} c_{i\alpha\sigma}+\sum_{\alpha \sigma} 
          \varepsilon^{a}_{\alpha}
          a^{\dag}_{i\alpha\sigma}a_{i\alpha\sigma}\nonumber\\
          &+&\sum_{\alpha \sigma}V_{\alpha}
          (c^\dag_{i\alpha\sigma}a_{i\alpha\sigma}+H.c.)+
{\cal H}_i',\label{eq_ref_model}
\end{eqnarray}
%--------------------------------------------------------------------
where $a^{\dag}_{i\alpha\sigma}(a_{i\alpha\sigma})$ 
creates (annihilates) an electron with spin $\sigma$ and orbital $\alpha$, 
which is connected to the $i$th site in the original lattice. 
The grand potential per site is rewritten as,
\begin{eqnarray}
\Omega /L&=&\Omega_{\rm imp}
    -2 \sum _{\alpha}\sum_m g( \omega'_{\alpha m}) 
-2\sum_\alpha g(\varepsilon^{a}_\alpha-\mu)\nonumber\\
    &+&2 \sum _{\alpha}\sum_m  \int^{-\infty}_{\infty}dz 
      \rho_\alpha(z) g[ \omega_{\alpha m}(z)],\label{eq_omg}\\
 g(x)&=&-T \ln [1+\exp(-x/T)]
\end{eqnarray}
%%%%%%%%%%%%%%%%%%%%%%%%%%%%%%%%%%%%%%%%%
where $\Omega_{\rm imp}$ is the grand potential for the reference system and
 $\omega_{\alpha m}(z)$ $[\omega'_{\alpha m}]$
are the poles of Green function $G$ $(G')$ for
the original (reference) system. 
The Green function of the reference system is given as,
\begin{eqnarray}
G'_{\alpha}(\omega)&=&[\omega+\mu-\varepsilon^{c}_{\alpha}
           -\frac{V_{\alpha}^2}{\omega+\mu-\varepsilon^{a}_{\alpha}}
-\Sigma_{\alpha}(\omega)]^{-1},
           %-\Delta_\alpha(\omega)-\Sigma_{\alpha}(\omega)]^{-1},
%G'_{k\alpha}(\omega)&=&(\omega+\mu-\varepsilon_{k\alpha})^{-1},\ 
%(k=1,2,\cdots,N_b),
\end{eqnarray}
where $\Sigma_\alpha(\omega)$ is the self-energy for the $\alpha$th orbital.
%$\Delta_\alpha(\omega)=V_{\alpha}^2/(\omega+\mu-\varepsilon^{a}_{\alpha})$.
On the other hand, the full Green function for the original system is given as,
\begin{eqnarray}
G_\alpha(\omega)&=&\int dz \rho_\alpha(z)G_\alpha(\omega;z),\\
G_\alpha(\omega;z)&=&[\omega+\mu-z-\Sigma_\alpha(\omega)]^{-1},
\end{eqnarray}
where $\rho_\alpha(x)$ is the density of states for the bare $\alpha$th orbital.
Note that the following differential equation is efficient to deduce 
the poles $\omega_{\alpha m}(z)$ of the Green function for the original system,
\begin{eqnarray}
\frac{d\omega_{\alpha m}(z)}{dz}=\left(1-\frac{\partial\Sigma_\alpha(\omega)}{
\partial \omega}\right)^{-1}_{\omega=\omega_{\alpha m}(z)}.
\end{eqnarray}
By solving these differential equations, 
we estimate the free energy [Eq. (\ref{eq_FreeEnergy})] numerically.
The variational parameters
${\boldsymbol \lambda}=(\mu, V_\alpha, \epsilon^c_\alpha, \epsilon^a_\alpha$)
are determined by the condition $\partial F/\partial {\boldsymbol \lambda}=0$,
which corresponds to the stationary point in the parameter space.

In the framework of SFA,
the variational parameters for the metallic and the OSM states can be
determined numerically.
In contrast,  
it may be difficult to determine unique variational parameters 
for the Mott insulating state, as discussed above.
To overcome this, we make use of the analytical and the numerical methods.

For simplicity, we consider the two-orbital model with $J/U=0$ 
at quarter filling.
Recall that 
the Mott insulating phase is simply represented in the atomic limit,
by taking into account local electron correlations.
Since the four one-electron states $c^\dag_{\alpha\sigma}|0\rangle$ 
are degenerate $(\varepsilon^c_1=\varepsilon^c_2)$,
the ground state $|\Phi_G\rangle$ in the reference system 
[eq. (\ref{eq_ref_model})]
should be given as
\begin{equation}
|\Phi_G\rangle=\sum_{\alpha\sigma} \sqrt{n_\alpha}c^\dag_{\alpha\sigma}|0\rangle,
\label{eq_ground_state}
\end{equation}
where %$|\sigma,\alpha\rangle=c^\dag_{\alpha\sigma}|0\rangle$,
$n_\alpha$ is the expectation value of band fillings 
for the $\alpha$th orbital, 
which satisfies the constraint $\sum_{\alpha\sigma} n_\alpha =1$.
%and $\varepsilon^c_1=\varepsilon^c_2$ is required for the degeneracy of 
%$|c^\dag_{\alpha\sigma}\rangle$.

The Green function and the self-energy for the reference system 
[eq. (\ref{eq_ref_model})] 
are given as 
\begin{eqnarray}
G^\prime_\alpha(\omega)&=&\frac{n_\alpha}{\omega+\mu+\varepsilon^c_\alpha}+
\frac{1-n_\alpha}{\omega+\mu+\varepsilon^c_\alpha-U},\\
\Sigma_\alpha(\omega)&=&
\omega+\mu+\varepsilon^c_\alpha-1/G^\prime_\alpha(\omega).
\end{eqnarray}
The Green function for the original system is then written as
$G_\alpha(\omega)=\int dz \rho_\alpha(z)G_\alpha(\omega;z)$
with
\begin{eqnarray}
G_\alpha(\omega;z)&=& \frac{1}{\omega+\mu-z-\Sigma_\alpha(\omega)}\nonumber\\
           &=& \sum_{l=\pm}\frac{R^\alpha_l(z)}{\omega-\xi_l^\alpha(z)},
\end{eqnarray}
where
\begin{eqnarray}
\xi_{\pm}^\alpha(z)&=&\frac{1}{2}\bigg(U+z-\varepsilon^c_\alpha \pm \eta(z)\bigg)
-\mu,\nonumber\\
R_{\pm}^\alpha(z)&=&\frac{1}{2}\bigg(1\pm
\frac{z+\varepsilon^c_\alpha+U(1-2n_\alpha)}
{\eta(z)}\bigg),\nonumber\\
\eta(z)&=&\sqrt{U^2+2(1-2n_\alpha)(z+\varepsilon^c_\alpha)U+(z+\varepsilon^c_\alpha)^2}.\nonumber
\end{eqnarray}
This formula is equivalent to the Hubbard-I type approximation.
Finally, we obtain the free energy per site for the system as
\begin{eqnarray}
F/L=\sum _\alpha(\mu+\varepsilon^c_\alpha)(1-n_\alpha)
+\mu\sum_\alpha n_\alpha\nonumber\\
+\sum_\alpha \sum_{l=\pm}\int dz\rho_\alpha(z) \xi_l^\alpha(z)\theta[-\xi^\alpha_l(z)].
\end{eqnarray}

At the stationary point in the free energy, three conditions are obtained as,
\begin{eqnarray}
1&=&\sum_{l=\pm}\int dz\rho_\alpha(z)\theta[-\xi^\alpha_l(z)],
\label{eq_condition_mu}\\
n_\alpha&=&\sum_{l=\pm}\int dz
\rho_\alpha(z)R_l^\alpha(z)\theta[-\xi^\alpha_l(z)], %(\alpha=1,2)\nonumber\\
\label{eq_condition_delta}
\end{eqnarray}
We wish to note that 
the right hand side of eq. (\ref{eq_condition_delta})
is equivalent to $\int^{i \infty}_{-i \infty} G_\alpha(\omega) d\omega/2\pi i$ 
for each orbital.
This implies that 
$n_\alpha$ is identical to the electron count for the lattice system.
%$n_\alpha=\int^{i \infty}_{-i \infty}G_\alpha(\omega) d\omega/2\pi i$ on
%the Green function $G_\alpha(\omega)$.

As described above, the chemical potential is difficult to be determined 
uniquely in the insulating phase.
In fact, the condition $\partial F/\partial \mu=0$ is satisfied 
in the case $W_\alpha/2\lesssim\mu\lesssim U-W_\alpha/2 \; (U >> W_\alpha)$, 
which originates from the presence of the Hubbard gap.
Here, %since we interest in a region $U>W_\alpha$,
we set the chemical potential as $\mu=U/2$. 
Then 
the condition eq. (\ref{eq_condition_mu}) is always satisfied, and
eq. (\ref{eq_condition_delta}) is rewritten as
\begin{equation}
n_\alpha=\int dz\rho_\alpha(z)R_-^\alpha(z). % \ \ (\alpha=1,2).
\label{eq_last_condition_2_3}
\end{equation}
By solving the self-consistent equations 
eq. (\ref{eq_last_condition_2_3}),
$\sum_{\alpha\sigma} n_\alpha = 1$, 
and $\varepsilon^c_1=\varepsilon^c_2$, 
we can describe the Mott insulating phase
in the framework of the SFA.

%\bibliography{main}

\end{document}